\documentclass{llncs}

\usepackage{amssymb}
\usepackage{subfigure}
\usepackage{color}
\usepackage{epsfig}
\usepackage{url}
\usepackage{multirow}

\begin{document}
\title{Full-text Support for Publish/Subscribe Ontology Systems}
\author{Lefteris Zervakis$^1$, Christos Tryfonopoulos$^1$, Antonios Papadakis-Pesaresi$^2$,\\ Manolis Koubarakis$^2$, and Spiros Skiadopoulos$^1$}
\institute{$^1$ Dept. of Computer Science and Technology, University of Peloponnese\\
$^2$ Dept. of Informatics and Telecommunications, University of Athens}
\maketitle
\vspace{-0.55cm}
\begin{abstract}
We envision a publish/subscribe ontology system that is able to index millions of user subscriptions and filter them against ontology data that arrive in a streaming fashion. In this work, we propose a \emph{SPARQL extension} appropriate for a publish/subscribe setting; our extension builds on the natural semantic graph matching of the language and supports the creation of \emph{full-text subscriptions}. Subsequently, we propose a main-memory \emph{subscription indexing algorithm} which performs both semantic and full-text matching at low complexity and minimal filtering time. Thus, when ontology data are published matching subscriptions are identified and notifications are forwarded to users.
\end{abstract}
\vspace{-0.9cm}
\section*{System overview}
\vspace{-0.3cm}

Resource Description Framework (RDF) constitutes a conceptual model and a formal language for representing resources in the Semantic Web. It is also the data format of choice for modern \emph{publish-subscribe} ontology systems, which demand sophisticated data representation and efficient filtering mechanisms to match massive ontology data against millions of \emph{user subscriptions} (also referred to as \emph{continuous queries}). The SPARQL query language is currently the W3C recommendation for querying RDF data and the Semantic Web. The graph model over which it operates naturally joins data together and represents a fully-fledged language; however, it lacks the support of a complete \emph{full-text retrieval} mechanism, beyond existing regular expression support, with sophisticated algorithms and data structures to minimise processing and memory requirements.

In this work, we focus on full-text filtering of ontology data that contain RDF literals in their property elements. To preserve the expressivity of SPARQL, we view the full text operations as an additional \emph{filter} of the subscription variables. In this context, we define a new binary operator $\mathit{ftcontains}$ that takes a variable of the subscription and a full-text expression that operates on the values of this variable as parameters. An example of a SPARQL subscription with full-text support is shown below.
    \begin{center}
    \scriptsize{
        \begin{tabular}{l l l l }
        $SELECT$ & $\ ?article$     &                  &           		\\
        $WHERE$  & $\{?publisher$ & $rdf:type$       & $Publisher.$		\\
                 & $\ ?publisher$ & $publishes$      & $?article.$     	   	\\
                 & $\ ?article$   & $articleText$    & $?articleText.$ 	   	\\
        $FILTER$ & $\ ftcontains$ & $(?articleText,$ & $``economic"\ ftand\ ``crisis")\}$\\
        \end{tabular}
    }
    \end{center}

We focus on RDF triples where the $subject$ is always a node element and the $predicate$ denotes the subject's relation to the $object$, which is a literal expressed as a typed or untyped string. A full text expression is evaluated only against a literal; thus the variable of the subscription can only be the object of a triple pattern.
The expressions supported involve the usual \emph{Boolean operators} (denoted by $\mathit{ftand}$, $\mathit{ftor}$, etc.), as well as \emph{proximity} and \emph{phrase} matching. Below we present an example of a full-text SPARQL subscription that will match all \emph{rdf:type Article} node elements, with a property named \emph{title} containing a string literal with the keywords \emph{``economic"} and \emph{``crisis"}.



\begin{figure}[t]
 \begin{minipage}[t]{0.5\linewidth}
    \centering
    \includegraphics[height=0.20\textheight, width=\linewidth]{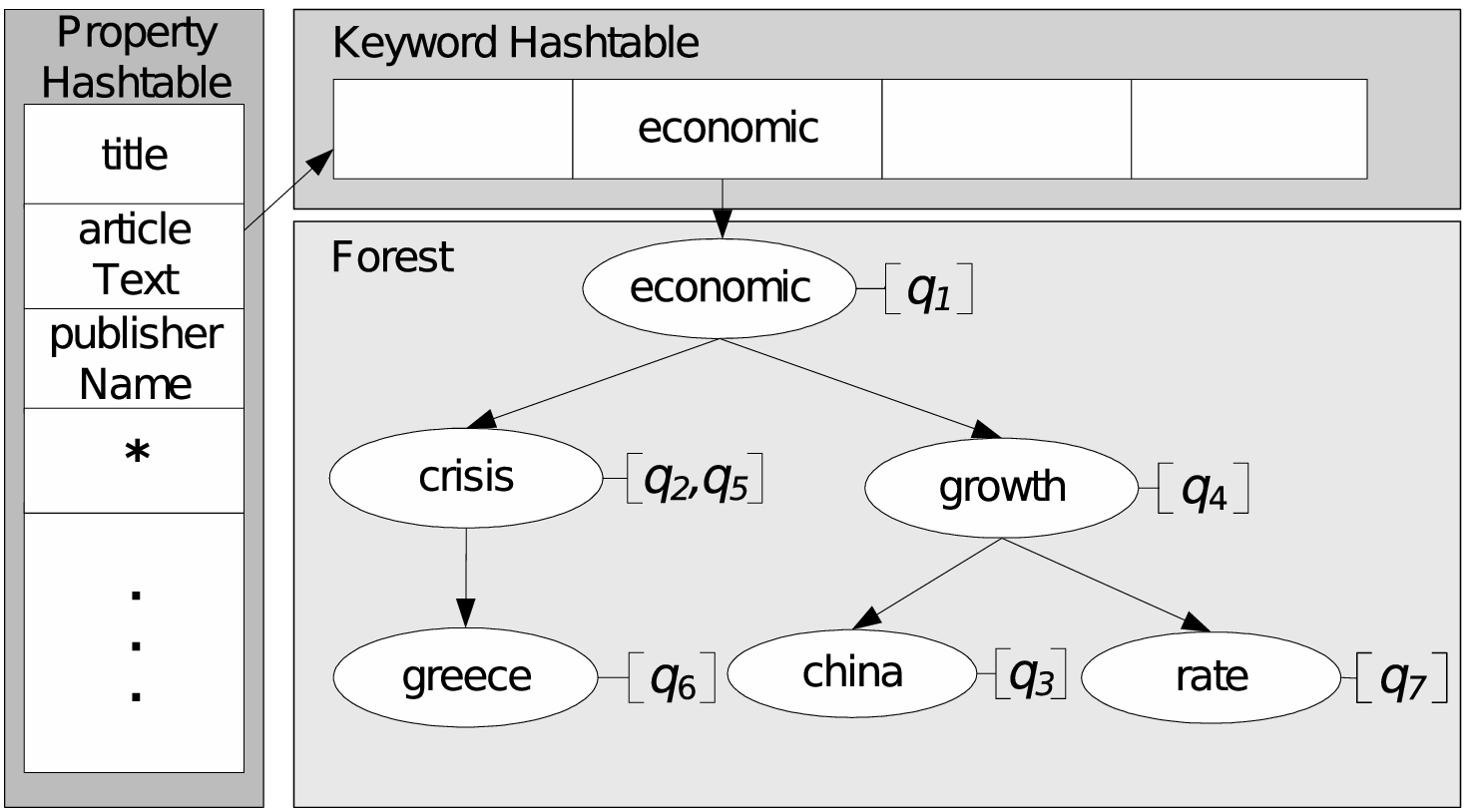}
    \caption{Subscription indexing scheme}
    \label{fig:forest}
  \end{minipage}
\begin{minipage}[t]{0.5\linewidth}
    \centering
    \includegraphics[height=0.20\textheight, width=\linewidth]{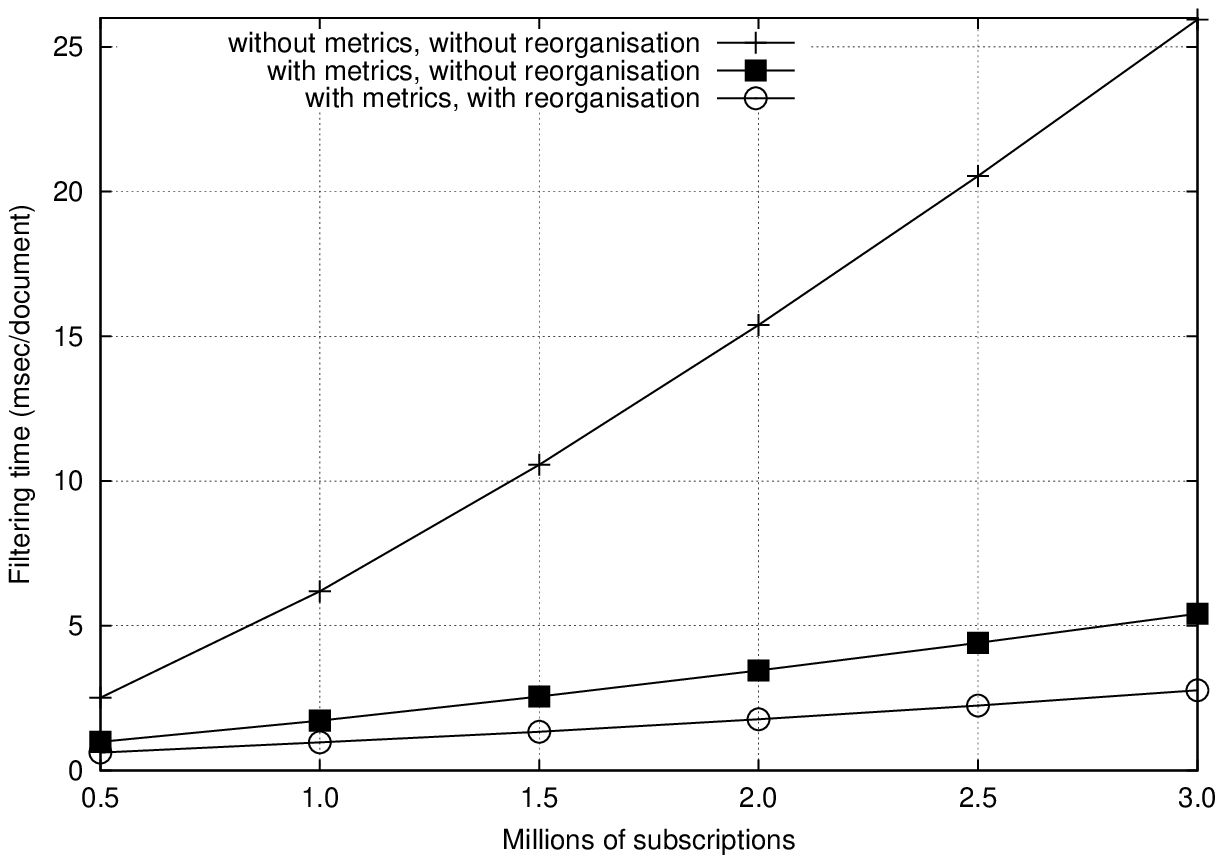}
    \caption{Filtering time/document (msecs)}
    \label{fig:time}
\end{minipage}
\vspace{-0.6cm}
\end{figure}

To perform the semantic matching, we define a Semantic Match Table in the spirit of \cite{ParkC09}, where a two-level hash table is used to represent the series of joins in a SPARQL subscription as a connected chain. We extend this idea to provide a hashing scheme that is able to accommodate all possible types of triple patterns in SPARQL subscriptions. Additionally, to support the full-text features introduced in the SPARQL subscriptions, we utilise a \emph{property hash table} that uses as key the constant part of the triple pattern in the SPARQL subscription. This hash table provides access to a data structure, which comprises of (i) \emph{tries} storing the keywords contained in the full-text part of subscriptions and (ii) a \emph{keyword hash table} that allows fast access to the trie roots.
Figure \ref{fig:forest} shows these data structures for a set of seven user subscriptions.

User subscriptions are organised into tries extending the approach of \cite{TryfonopoulosKD09} to rely on \emph{common subsets} of subscriptions. The main idea behind the indexing algorithm is to use tries to capture common elements of subscriptions. To do so, we utilise \emph{metrics} to locate the best possible indexing position in the forest of tries. Since our algorithm is influenced by the order of insertion of subscriptions (due to greedy subscription indexing), a statistics-based \emph{subscription reorganisation} is employed. In the reorganisation phase of the algorithm, a scoring mechanism is utilised to modify the order of subscription indexing for all subscriptions inserted since the last reorganisation of the forest.
In our evaluation we used $3.1M$ extended abstracts downloaded from DBpedia as incoming RDF documents and artificially generated subscription databases of varying sizes. Figure \ref{fig:time} shows the filtering time when (i) no metrics for the best indexing position in the forest are employed (deterministic subscription indexing), (ii) metrics are employed, but no re-organisation is used, and (iii) both metrics and reorganisation are employed.
\vspace{-0.5cm} 
\bibliographystyle{splncs03}
\scriptsize{
\bibliography{eswc}}

\begin{thebibliography}{1}
\providecommand{\url}[1]{\texttt{#1}}
\providecommand{\urlprefix}{URL }
\vspace{-0.3cm}

\bibitem{ParkC09}
Park, M.J., Chung, C.W.: ibroker: An intelligent broker for ontology based
  publish/subscribe systems. In ICDE 2009.

\bibitem{TryfonopoulosKD09}
Tryfonopoulos, C., Koubarakis, M., Drougas, Y.: Information filtering and query
  indexing for an information retrieval model. In ACM TOIS 2009.

\end{thebibliography}
\end{document}